\documentclass[aps,pre,superscriptaddress,showkeys,citeautoscript]{revtex4-2}
\usepackage[colorlinks=true,bookmarks=false,citecolor=blue,linkcolor=red,hyperfootnotes=true,urlcolor=blue]{hyperref}

\usepackage{physics}
\usepackage{graphicx}% Include figure files
\usepackage{dcolumn}% Align table columns on decimal point
\usepackage{bm}% bold math
\usepackage[dvipsnames]{xcolor}
\usepackage{wasysym}
\usepackage{lipsum}
\usepackage{enumitem}
\usepackage{booktabs} 
\usepackage{gensymb}
\usepackage{xcolor}
\usepackage{amssymb}
\usepackage{booktabs}
\usepackage{array}
\newcolumntype{L}{>{$}l<{$}}
\newcolumntype{C}{>{$}c<{$}}
\newcolumntype{R}{>{$}r<{$}}

\usepackage{blindtext}

\makeatletter
\def\p@subsection{}
\makeatother

\newcommand{\code}[1]{\texttt{#1}}

\begin{document}

\title{The Generalized Green's function Cluster Expansion: A Python package for simulating polarons}

\author{Matthew R. Carbone}\email{mcarbone@bnl.gov}
\thanks{Equally-contributing author}
\affiliation{Computational Science Initiative, Brookhaven National Laboratory, Upton, New York 11973, USA}

\author{Stepan Fomichev}
\thanks{Equally-contributing author}
\affiliation{Department of Physics and Astronomy, University of British Columbia, Vancouver, British Columbia V6T 1Z1, Canada}
\affiliation{Stewart Blusson Quantum Matter Institute, University of British Columbia, Vancouver, British Columbia, V6T 1Z4 Canada}

\author{Andrew J. Millis}
\affiliation{Department of Physics, Columbia University, New York, New York 10027, USA}
\affiliation{Center for Computational Quantum Physics, Flatiron Institute, New York, New York 10010, USA}

\author{Mona Berciu}
\affiliation{Department of Physics and Astronomy, University of British Columbia, Vancouver, British Columbia V6T 1Z1, Canada}
\affiliation{Stewart Blusson Quantum Matter Institute, University of British Columbia, Vancouver, British Columbia, V6T 1Z4 Canada}

\author{David R. Reichman}
\affiliation{Department of Chemistry, Columbia University, New York,
New York 10027, USA}

\author{John Sous}\email{sous@stanford.edu}
\affiliation{Department of Physics, Stanford University, Stanford, California 93405, USA}
\affiliation{Geballe Laboratory for Advanced Materials, Stanford University, Stanford, California 94305, USA}

\date{\today}

\begin{abstract}
We present an efficient implementation of the Generalized Green's function Cluster Expansion (GGCE), which is a new method for computing the ground-state properties and dynamics of polarons (single electrons coupled to lattice vibrations) in model electron-phonon systems. The GGCE works at arbitrary temperature and is well suited for a variety of electron-phonon couplings, including, but not limited to, site and bond Holstein and Peierls (Su-Schrieffer-Heeger) couplings, and couplings to multiple phonon modes with different energy scales and coupling strengths. Quick calculations can be performed efficiently on a laptop using solvers from NumPy and SciPy, or in parallel at scale using the PETSc sparse linear solver engine.
\end{abstract}

\keywords{}

\pacs{}
\maketitle

\section{Statement of need}
The electron-phonon problem is of both fundamental relevance and practical importance in materials science~\cite{Mahan,ephReview2}. Electron-phonon interactions promote a variety of states, low-temperature phases and high-temperature transport phenomena in quantum materials. For example, they are essential to the understanding of the behavior of solar cells~\cite{solarcells} and semiconductors~\cite{semiconductors}. In the dilute-carrier-density limit, electron-phonon coupling gives rise to quasiparticles called polarons whose properties encode the physics of materials in various temperature regimes.

Research on polarons has been divided into two thrusts: fundamental theoretical work focused on qualitative physical aspects~\cite{Mahan} and applied research focused on obtaining quantitative properties relevant to specific materials~\cite{giustino2017electron,sio2019polarons,sio2019ab,ponce2016epw,zhou2016ab,zhou2021perturbo,lee2018charge}. This paper presents a new scientific software that aims to bridge this gap. It allows for the treatment of polaron statics and dynamics in models of electron-phonon coupling of almost arbitrary form provided that they are sufficiently short-ranged. It presents a self-contained first step in an ongoing effort to combine an \textit{ab initio} understanding of materials and exact many-body analysis of polaron states.

\section{Software summary}

The GGCE method is a \textit{numerically exact} extension of a family of variational approaches known in the theoretical physics community as Momentum Average (MA) methods~\cite{Berciu2006prl,Goodvin2006prb}. Details on the theoretical framework of GGCE can be found in Carbone \textit{et al}~\cite{carbone2021numerically,carbone2021bond}. Our code, named for the method, is a Python package meant to make implementing the GGCE framework as straightforward as possible. In addition, through only slight modifications to our standard API, the user can invoke powerful PETSc sparse solvers for massively parallel computations at scale~\cite{petsc-web-page,petsc-user-ref,petsc-efficient,PetscSF2022}.

A fundamental insight of the MA approximation is to utilize a variational space formed of clouds of spatially clustered phonon configurations with the electron allowed to be anywhere in the system. Through comparison with exact methods, this approximation was shown to yield quantitatively accurate results. In order to systematically converge MA to the limit of infinite Hilbert space dimension, the cloud size and total phonon number, which serve as control parameters, are taken to infinity~\cite{Goodvin2006prb}. This, however, requires derivation of a set of equations corresponding to a given cloud size for all cloud sizes smaller than a cutoff. This cutoff is then increased until convergence is achieved. The ever-increasing complexity of the structure of the system of equations at large cloud sizes means that this approach can very quickly become intractable to do by hand~\cite{Dominic,Sous1,Sous2}, especially in the regime of small phonon energies where large cloud sizes are usually needed in order to converge to the numerically exact limit. Carbone \textit{et al} proposed a a generalized implementation of the MA method which automates the generation and solution of the systems of equations for arbitrary cloud sizes~\cite{carbone2021numerically}. Benchmarks of GGCE on several model systems verified that convergence with cloud size is fast, rendering this an efficient and controlled numerically exact method even in challenging parameter regimes. 

Previous studies using GGCE focused on polarons at zero temperature. We also include a new functionality which allows the study of polarons at finite temperature. Here, we make use of the Thermofield Double formalism~\cite{suzuki1985thermo,takahashi1996thermo}, which exactly maps any given model at finite temperature to one at zero temperature with couplings to real and fictitious phonons. This model can be solved naturally using the apparatus of the zero-temperature GGCE method. Benchmarks of this approach are ongoing, and preliminary results suggest that the method may be competitive with state-of-the-art methods. A paper with these results is currently in preparation.

Formally, the GGCE functions as an on-the-fly generator of equations of motion for the single-particle Green's function given a set of control parameters (cloud size, total phonon number, and their extensions to systems with multiple phonon modes) and input model parameters (energy scales, coupling strength, etc.). The generated system of equations is then solved numerically in order to obtain the Green's function of interest using a chosen solver. Furthermore, the equation of motion dictates how different Green's functions or propagators couple, and so one can use the solver to numerically obtain any one particular propagator or a set of them, which can be used to construct other quantities such as the optical conductivity~\cite{optical} or resonant inelastic X-ray scattering spectrum~\cite{Rixs}.

The GGCE code consists of three components detailed in our documentation: models, systems and solvers.
\begin{itemize}
    \item Models completely describe the Hamiltonian system to solve, and the level of theory (specified by the control parameters) at which to solve it;
    \item Systems construct all of the objects required to build the matrix to solve the system of equations;
    \item Solvers utilize different back-ends to actually solve the constructed matrix in an efficient manner.
\end{itemize}

\subsection{Models}

The choice of model completely defines the type of electron-phonon coupling used in the Hamiltonian. Every model Hamiltonian implemented so far assumes a lattice with nearest-neighbor hopping of electrons and Einstein (dispersionless) phonons. The user can set the electron hopping, phonon frequency and type and strength of the electron-phonon coupling. Currently, we have implemented the Holstein, site Peierls (site Su-Schrieffer-Heeger), bond Peierls (bond Su-Schrieffer-Heeger) and Edwards Fermion-boson models (as well as any arbitrary combination of these).

\subsection{Systems}

The Systems objects are a helpful intermediary for performing the sometimes expensive step of constructing the Python objects required for building matrices of linear equations. Systems are instantiated from a Model. At creation, using the information in the Model about the electron-phonon couplings, Systems automatically construct and store the equations-of-motion object called the ``basis". The basis remains ``un-evaluated": it is passed into the Solver, where it can be used to obtain a matrix of equation coefficients at any values of momentum and frequency. In this way, the basis is constructed only once, and then simply called repeatedly to determine the coefficients. Construction of the basis follows the scheme outlined in Carbone \textit{et al}~\cite{carbone2021numerically}. For relatively small clouds, the basis can be visualized for sanity checking the calculation, using a method that pretty-prints the equations' structure.

When provided a directory, the System will also automatically checkpoint the basis to disk using \code{pickle}, allowing for a later restart of a failed computation or for restarting long jobs on a cluster with time-limited jobs without the expensive re-computation of the basis.

\subsection{Solvers}

At the heart of GGCE are the Solvers, which implement different approaches to solving the linear systems of equations obtained by a System object. The simplest of these uses NumPy's dense solver, which solves the equation-of-motion matrix using an efficient continued fraction approach~\cite{Goodvin2006prb}, or SciPy's sparse solver. For truly large-scale computations with sizable phonon clouds and/or many different electron-phonon couplings operating simultaneously, GGCE interfaces with the powerful, massively parallel PESTc sparse solver engine. All GGCE Solvers are MPI-enabled, and allow for a variety of parallelization schemes, all of which are detailed in our documentation. At the extreme, the PETSc interface can parallelize calculations across momentum-frequency points and also parallelize the solving of a single large sparse matrix at each point, and thus allowing for straightforward use of all available computational resources.

The Solver allows the user to quickly evaluate the Green's function for a specified range of momenta and frequencies. Like the System, it automatically checkpoints the solution (Green's function value) at every momentum-frequency point using pickle, allowing for restart in case of failure or time limits on cluster jobs.

\section{Concluding notes}
The GGCE package can be found open source under a BSD-3-clause license at \href{https://github.com/x94carbone/GGCE}{github.com/x94carbone/GGCE}, or can be installed via pip using \code{pip install ggce}.

\begin{acknowledgements}
M. R. C. acknowledges the following support: This material is based upon work supported by the U.S. Department of Energy, Office of Science, Office of Advanced Scientific Computing Research, Department of Energy Computational Science Graduate Fellowship under Award Number DE-FG02-97ER25308. S. F. and M. B. acknowledge support from the Natural Sciences and Engineering Research Council of Canada (NSERC) and the Stewart Blusson Quantum Matter Institute (SBQMI). A. J. M., D. R. R. and J. S. acknowledge support from the National Science Foundation (NSF) Materials Research Science and Engineering Centers (MRSEC) Program through Columbia University in the Center for Precision Assembly of Superstratic and Superatomic Solids under Grant No. DMR-1420634. J. S. acknowledges support from the Gordon and Betty Moore Foundation's EPiQS Initiative through Grant GBMF8686 at Stanford University. The Flatiron Institute is a division of the Simons Foundation.

Disclaimer: This report was prepared as an account of work sponsored by an agency of the United States Government. Neither the United States Government nor any agency thereof, nor any of their employees, makes any warranty, express or implied, or assumes any legal liability or responsibility for the accuracy, completeness, or usefulness of any information, apparatus, product, or process disclosed, or represents that its use would not infringe privately owned rights. Reference herein to any specific commercial product, process, or service by trade name, trademark, manufacturer, or otherwise does not necessarily constitute or imply its endorsement, recommendation, or favoring by the United States Government or any agency thereof. The views and opinions of authors expressed herein do not necessarily state or reflect those of the United States Government or any agency thereof.
\end{acknowledgements}

\providecommand{\noopsort}[1]{}\providecommand{\singleletter}[1]{#1}%

\end{document}